\documentclass[%
 reprint,
 superscriptaddress,
 preprintnumbers,
 nofootinbib,
 amsmath,amssymb,
 aps,
 prl,
]{revtex4-1}
\usepackage{psfrag,slashed,cancel,array,graphicx,todonotes,hyperref}
\usepackage{subfigure}
\usepackage[labelfont=bf,labelsep=quad,justification=justified]{caption}
\usepackage[utf8]{inputenc}
\usepackage{mathtools}
\usepackage{mathrsfs}
\usepackage{multirow}
\usepackage{booktabs} 
\usepackage[normalem]{ulem}
\hypersetup{pdftitle={},pdfcreator={},linkcolor=[rgb]{0.15,0.35,0.75},colorlinks=true,citecolor=[rgb]{0.675,0,0.2},urlcolor=[rgb]{0.15,0.35,0.65}}

\def\beq{\begin{equation}}
\def\eeq{\end{equation}}
\def\bsp#1\esp{\begin{split}#1\end{split}}

\AtBeginDocument{
\heavyrulewidth=.08em
\lightrulewidth=.05em
\cmidrulewidth=.03em
\belowrulesep=.65ex
\belowbottomsep=0pt
\aboverulesep=.4ex
\abovetopsep=0pt
\cmidrulesep=\doublerulesep
\cmidrulekern=.5em
\defaultaddspace=.5em
}

\def\eps{\epsilon}

\begin{document}

\preprint{CERN-TH-2020-009}
\preprint{MIT-CTP/5174}
\preprint{SLAC-PUB-17508}

\title{The Drell-Yan cross section to third order in the strong coupling constant}

\author{Claude Duhr}
\affiliation{Theoretical Physics Department, CERN,
CH-1211 Geneva 23, Switzerland.}
\email{claude.duhr@cern.ch}
\author{Falko Dulat}
\affiliation{SLAC National Accelerator Laboratory, Stanford University, Stanford, CA 94039, USA.}
\email{falko.dulat@gmail.com}
\author{Bernhard Mistlberger}
\affiliation{Center for Theoretical Physics, Massachusetts Institute of Technology, Cambridge, MA 02139, USA.}
\email{bernhard.mistlberger@gmail.com}
\begin{abstract}
We present phenomenological results for the inclusive cross section for the production of a lepton-pair via virtual photon exchange at next-to-next-to-next-to-leading order (N$^3$LO) in perturbative QCD. 
In line with the case of Higgs production, we find that the hadronic cross section receives corrections at the percent level, and the residual dependence on the perturbative scales is reduced. 
However, unlike in the Higgs case, we observe that the uncertainty band derived from scale variation is no longer contained in the band of the previous order.
\end{abstract}

\maketitle

The discovery of the Higgs boson at the Large Hadron Collider (LHC) at CERN~\cite{Aad:2012tfa,Chatrchyan:2012xdj} and the absence of signals of new physics has resulted in a new precision collider program. 
This program was made possible by parallel advances on both the theoretical and experimental sides. 
On the experimental side, we have seen the use of advanced experimental techniques and an improved and solid understanding of systematic uncertainties. 
On the theoretical side, we have lived through a revolution in our ability to perform higher-order computations for collider observables. 
These two things combined have made the LHC a precision machine, where measurements at the percent level will be achieved routinely and compared to theoretical predictions at a similar level of accuracy.

The Drell-Yan (DY) process is the shining example of the precision phenomenology program at the LHC.
This process  corresponds to the production of a pair of charged leptons with a fixed invariant mass $Q^2$ from an off-shell photon or $Z$-boson in Quantum Chromodynamics (QCD).
Its clean final-state signature makes it an ideal candidate for luminosity measurements and detector calibration. 
Moreover, the DY process plays a key role in the measurement of parton distribution functions (PDFs) at the LHC, and also in many searches for physics beyond the Standard Model (SM). 
Its importance for the physics program at the LHC makes the DY process one of the main processes for which very precise theoretical predictions are desirable. 
In particular, we need to have a clear understanding of its perturbative stability and of the size of the theoretical uncertainties.

The LHC being a hadron collider, one expects the most important perturbative
corrections to arise due to QCD\@.
The next-to-leading order (NLO) QCD corrections to lepton-pair production were computed already four decades ago in ref.~\cite{Altarelli:1978id,Altarelli:1979ub}. 
The next-to-next-to-leading order (NNLO) corrections were computed in ref.~\cite{Matsuura:1987wt,Matsuura:1988nd,Matsuura:1988sm,Matsuura:1990ba,Hamberg:1990np,vanNeerven:1991gh,Harlander:2002wh}, and were supplemented by electroweak (EW) corrections in ref.~\cite{Baur:2001ze,Baur:1997wa}. 
Very recently, also mixed QCD-EW corrections have become available~\cite{Delto:2019ewv,Bonciani:2019nuy}.

In the remainder of this letter, we focus on corrections in the strong coupling constant. 
The known results at NLO and NNLO indicated a good convergence of the perturbative series.
In particular, the size of the missing higher-order terms was estimated be at
the percent level at NNLO\@.
This estimate was obtained by varying the renormalisation and factorisation scales by a factor of two around the hard scale of the process set by the invariant mass of the lepton pair. 
Given the phenomenological importance of the DY process for the precision physics program at the LHC, it is important to have reliable predictions for the DY process and the associated theoretical uncertainties. 
In this letter we compute for the first time the next-to-next-to-next-to-leading order (N$^3$LO) corrections to lepton-pair production from a virtual photon. 
While the complete description of the DY process at this order will also require the contribution from the off-shell $Z$-boson (as well as the interference between the photon and the $Z$-boson), the contribution from virtual photon production already gives valuable information about the size of the QCD corrections and the convergence of the perturbative series. 
Specifically, we expect the relative size of perturbative corrections for $Z$ boson and virtual photon exchange to be very similar. 
We have checked this explicitly at lower orders in perturbation theory.
In addition, this is motivated by the fact that the analytic partonic cross sections are largely the same (up to differences related to diagrams with an axial vector current coupling to a single fermion trace starting at NNLO).
Furthermore, we found in ref.~\cite{Duhr:2020sdp} that the size of Drell-Yan K-factors is independent of the type of the electro-weak gauge boson through N$^3$LO.
Moreover, for small values of the invariant mass $Q^2$, the value of the cross section is dominated by the photon contribution, so that our results will provide reliable estimates of the size of the N$^3$LO corrections in that region. 
In the remainder of this letter, we review the computation of the N$^3$LO corrections to the photon contribution to the DY process and discuss its phenomenological implications.

\begin{figure*}[!t]
\setlength{\abovecaptionskip}{17.5 pt plus 4pt minus 2pt}
\mbox{
\subfigure{\includegraphics[scale=.23]{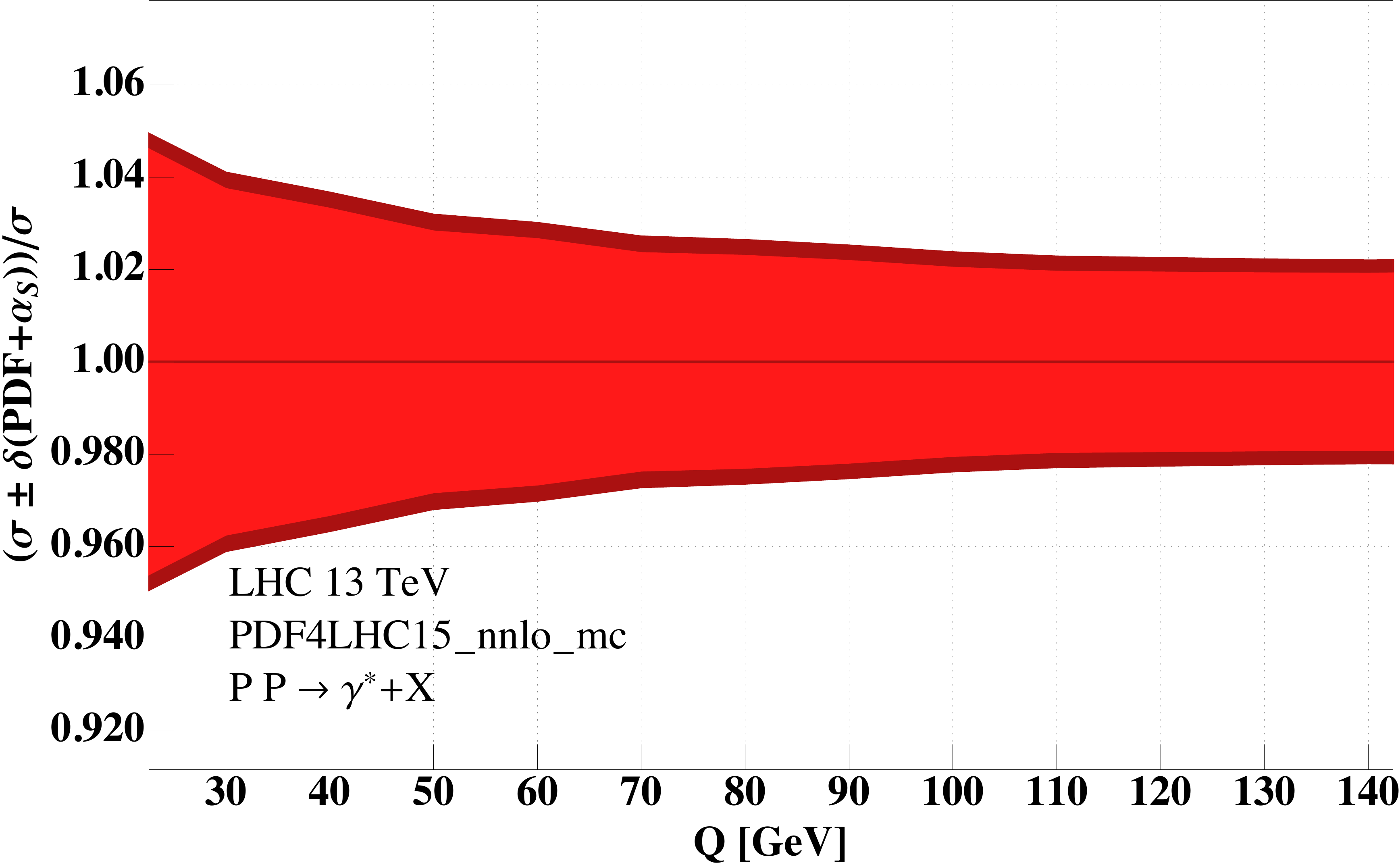}}\hskip 1cm
\subfigure{\includegraphics[scale=.23]{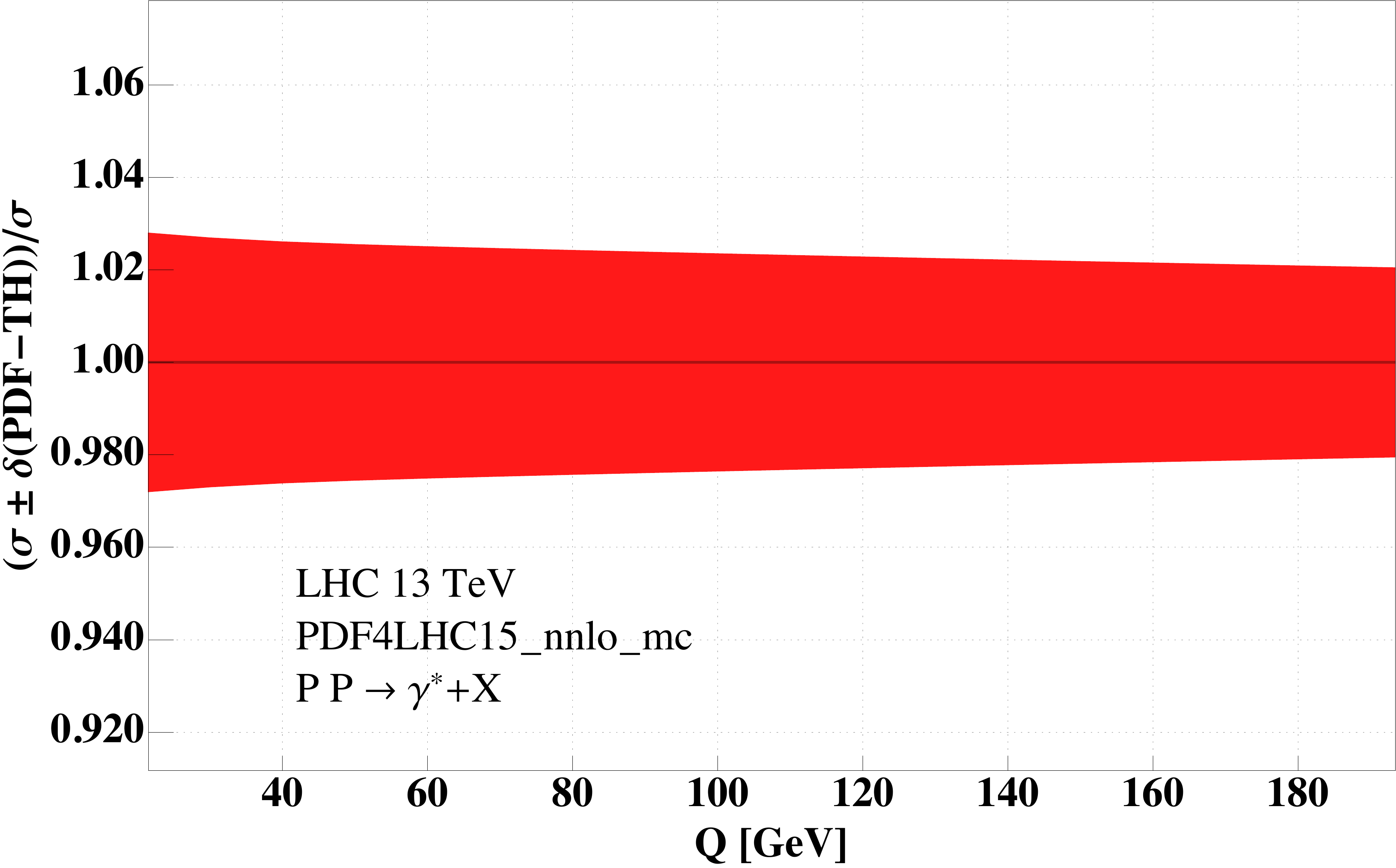}}
}
\vspace{-0.6 cm}
\caption{\label{fig:pdfs}
The light red area in the left plot represents the PDF uncertainty, the dark red area corresponds to the combination in quadrature of PDF$+\alpha_s$ uncertainty.
The right plot shows  the uncertainty on the cross section due to missing N$^3$LO PDFs.
}
\end{figure*}

\section{The N$^3$LO Corrections to the Drell-Yan Process}
The inclusive cross section for the production of a lepton-pair with invariant mass $Q^2$ can be written as
\beq
\label{eq:fac}
\frac{d\sigma}{dQ^2} \!=\! \sum_{i,j}\!\int_0^1\!\! dx_1\,dx_2\,f_i(x_1,\mu_f)f_j(x_2,\mu_f)\hat{\sigma}_{ij}(z,\mu_r,\mu_f)\,,
\eeq
where the sum runs over all parton flavours, $f_i$ are parton densities and $\hat{\sigma}_{ij}$ are partonic cross sections. 
The partonic cross sections depend on the ratio $z=Q^2/s$, where $\sqrt{s}$ is the partonic centre-of-mass energy, related to the hadronic centre-of-mass energy $\sqrt{S}$ by $s=x_1x_2S$ through the two Bjorken momentum fractions $x_{1,2}$. $\mu_r$ and $\mu_f$ denote the renormalisation and factorisation scales respectively. 
We have computed the partonic cross sections analytically through N$^3$LO for all partonic channels. 
At NLO and NNLO we reproduce the results of refs.~\cite{Altarelli:1978id,Altarelli:1979ub,Matsuura:1987wt,Matsuura:1988nd,Matsuura:1988sm,Matsuura:1990ba,Hamberg:1990np,vanNeerven:1991gh,Harlander:2002wh}. 
Our computation follows closely the one for the inclusive cross sections for Higgs production in gluon-fusion~\cite{Anastasiou:2015ema,Anastasiou:2016cez,Mistlberger:2018etf} and bottom-quark fusion~\cite{Duhr:2019kwi}.
All relevant Feynman diagrams are generated with {\sc QGraf}~\cite{qgraf} and sorted into scalar integral topologies, which are then reduced to a set of master integrals via integration-by-parts identities~\cite{Chetyrkin:1981qh,Tkachov:1981wb} using an in-house code. 
The master integrals are computed analytically as a function of $z$ using the differential equations method~\cite{Kotikov:1990kg,Kotikov:1991hm,Kotikov:1991pm,Henn:2013pwa,Gehrmann:1999as}. 
The master integrals contributing to the N$^3$LO cross section can be subdivided into several classes. 
Firstly, there are purely virtual three-loop integrals, which are encoded in the quark form factor up to three loops~\cite{Gehrmann:2006wg,Heinrich:2007at,Heinrich:2009be,Lee:2010cga,Baikov:2009bg,Gehrmann:2010ue,Gehrmann:2010tu}. 
We have recomputed the purely virtual corrections, and we find perfect agreement with the existing results in the literature. 
The N$^3$LO cross section also receives contributions from partonic subprocesses describing additional final-state radiation. 
The master integrals describing the emission of a single massless parton at this order in perturbation theory have been computed in ref.~\cite{Anastasiou:2013mca,Kilgore:2013gba,Duhr:2013msa,Li:2013lsa,Dulat:2014mda}.
Similarly, the master integrals for double-real virtual and triple-real contributions have been computed in refs.~\cite{Anastasiou:2014vaa,Li:2014bfa,Li:2014afw,Anastasiou:2015yha,Anastasiou:2013srw,Anastasiou:2015ema} as an expansion around the production threshold of the Higgs boson and exactly as a function of $z$ in ref.~\cite{Mistlberger:2018etf}. 
We work exclusively with the master integrals of ref.~\cite{Mistlberger:2018etf}. 
All master integrals have already been evaluated in the context of the N$^3$LO corrections to the gluon-fusion and bottom-quark-fusion cross sections.

The different contributions that we have described are not yet well-defined in four space-time dimensions. 
They are individually ultraviolet (UV) and infrared (IR) divergent, and we regulate both UV and IR using conventional dimensional regularisation, i.e., we work in $D=4-2\eps$ space-time dimensions. 
The UV divergences are absorbed by replacing the strong coupling constant by its renormalised value in the $\overline{\textrm{MS}}$-scheme. 
The UV-counterterm for the strong coupling constant has been computed through five loops in refs.~\cite{Tarasov:1980au,Larin:1993tp,vanRitbergen:1997va,Baikov:2016tgj,Herzog:2017ohr}. 
After UV renormalisation, all remaining divergences are of IR origin. 
They can be absorbed into the definition of the PDFs using mass factorisation at N$^3$LO~\cite{Buehler:2013fha,Hoschele:2012xc,Hoeschele:2013gga}, which involves convoluting lower-order partonic cross sections with the three-loop splitting functions of refs.~\cite{Moch:2004pa,Vogt:2004mw,Ablinger:2017tan}. 
All convolutions are computed analytically in $z$ space using the {\sc PolyLogTools} package~\cite{Duhr:2019tlz}. 
We observe that after UV renormalisation and mass factorisation, all poles in the dimensional regulator cancel and we obtain finite results for all partonic channels.

 Besides the explicit analytic cancellation of the UV and IR poles, we have performed various checks to establish the correctness of our computation.
 First, we have reproduced the soft-virtual N$^3$LO cross section of refs.~\cite{Eynck:2003fn,Moch:2005ky,Ahmed:2014cla,Li:2014bfa,Catani:2014uta} and the physical kernel constraints of ref.~\cite{Moch:2009hr,Soar:2009yh,deFlorian:2014vta} for the next-to-soft term of the quark-initiated cross section. 
 Second, we have checked that our partonic cross sections have the expected behaviour in the high-energy limit, which corresponds to $z\to 0$~\cite{MarzaniPhD,Marzani:2008uh}.
 Finally, we have also checked that all logarithmic terms in the renormalisation and factorisation scales produced from the cancellation of the UV and IR poles satisfy the Dokshitzer-Gribov-Lipatov-Altarelli-Parisi (DGLAP) evolution equation~\cite{Gribov:1972ri,Altarelli:1977zs,Dokshitzer:1977sg}.

  \begin{figure*}[!t]
\setlength{\abovecaptionskip}{17.5 pt plus 4pt minus 2pt}
\mbox{
   \includegraphics[scale=0.23]{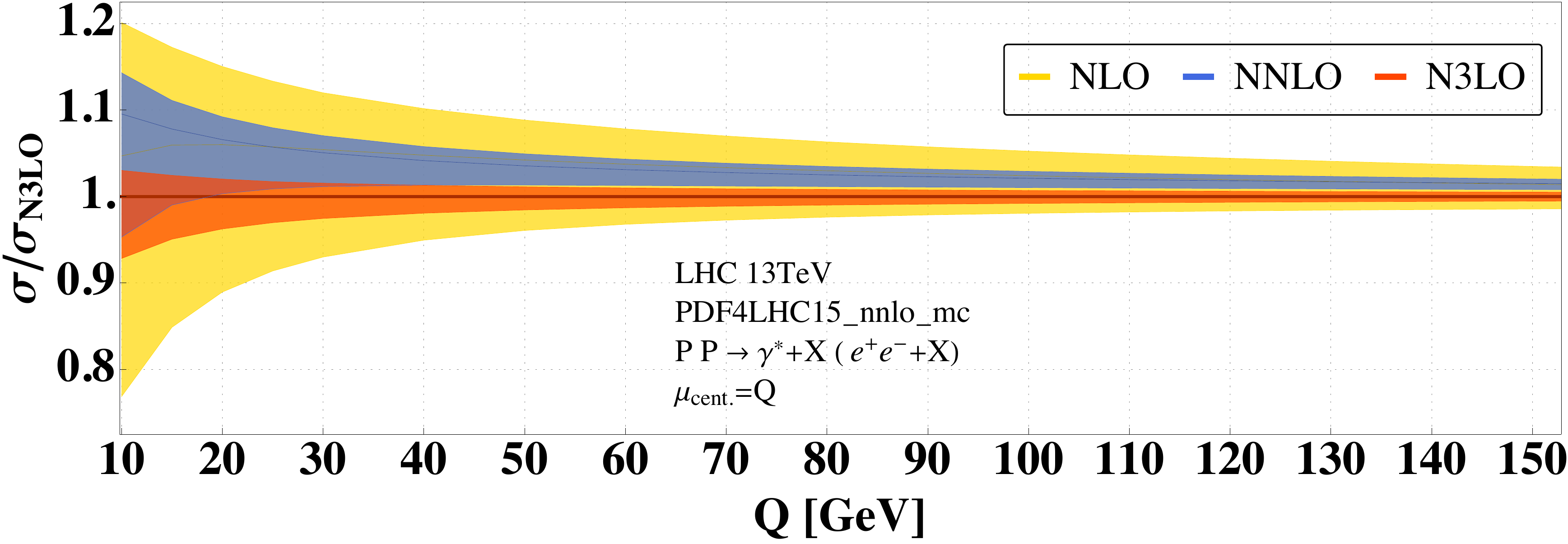}
   \hspace{0.5cm}
   \includegraphics[scale=0.23]{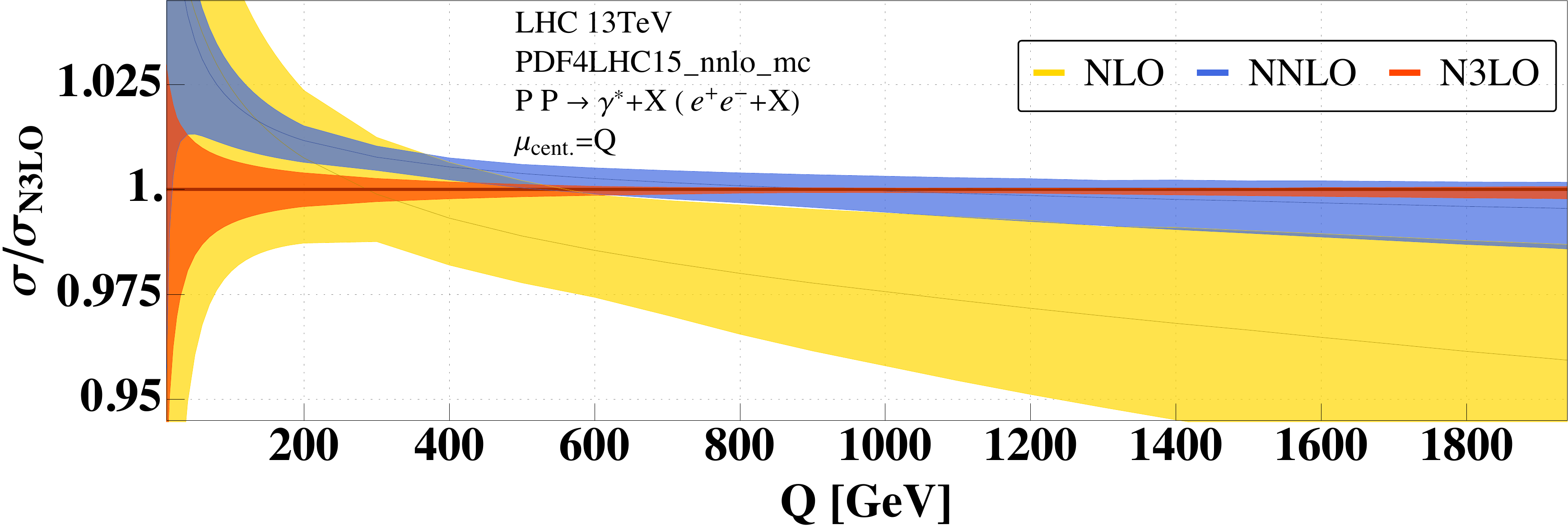}
}
\vspace{-0.6 cm}
 \caption{\label{fig:Q_variation}The cross section as a function of the invariant mass $Q^2$ of the lepton pair for small (left) and large (right) values of $Q$.}
\end{figure*}

  \section{Phenomenological Results}

In this section we present our phenomenological results for lepton-pair production via an off-shell photon at N$^3$LO in QCD. 
The strong coupling is $\alpha_s(m_Z^2) = 0.118$, and we evolve it to the renormalisation scale $\mu_r$ using the four-loop QCD beta function in the $\overline{\textrm{MS}}$-scheme assuming $N_f=5$ active, massless quark flavours. In the remainder of this section we present our results for the cross section as a function of the invariant mass of the lepton pair, and we discuss the sources of uncertainty that affect it. 

\begin{table}[!h]
\begin{equation}
\begin{array}{cccccc}\hline
    Q/\textrm{GeV} & \textrm{K}_{\text{QCD}}^{\text{N$^3$LO}} & \delta(\text{scale}) & \delta(\text{PDF+}\alpha_S) & \delta(\text{PDF-TH}) & \frac{\sigma_{Z+\gamma^*}^{(0)}}{\sigma_{\gamma^*}^{(0)}}\\
\hline
  30 & 0.952  & {}^{+1.5\%}_{-2.5 \%}  &  \pm 4.1 \%  &  \pm 2.7\% & 1.01 \\ \hline  
  50 & 0.966  & {}^{+1.1\%}_{-1.6 \%}  &  \pm 3.2 \%  &  \pm 2.5\% & 1.09\\ \hline  
  70 & 0.973  & {}^{+0.89\%}_{-1.1 \%}  &  \pm 2.7 \%  &  \pm 2.4\% & 2.16 \\ \hline  
  90 & 0.978  & {}^{+0.75\%}_{-0.89 \%}  &  \pm 2.5 \%  &  \pm 2.4\% & 415 \\ \hline  
  110 & 0.981  & {}^{+0.65\%}_{-0.73 \%}  &  \pm 2.3 \%  &  \pm 2.3\% & 7.4 \\ \hline  
  130 & 0.983  & {}^{+0.57\%}_{-0.63 \%}  &  \pm 2.2 \%  &  \pm 2.2\% & 3.5 \\ \hline  
  150 & 0.985  & {}^{+0.50\%}_{-0.54 \%}  &  \pm 2.2 \%  &  \pm 2.2\% & 2.6 \\ \hline  
\end{array}
\nonumber
\end{equation}
\caption{\label{tab:values} Numerical predictions for the QCD K-factor at N$^3$LO.}
\end{table}
Tab.~\ref{tab:values}  contains numerical values for the QCD K-factor, i.e., the ratio of the N$^3$LO cross section over the NNLO cross section. 
We observe that for all values of the invariant mass $Q$ considered, the cross section receives negative corrections at the percent level at LHC center-of-mass energies.
We include numerical estimates of the size of the three uncertainties discussed.
The central values and scale variation bands for the K-factor are obtained with the zeroth member of the \texttt{PDF4LHC15\_nnlo\_mc} set.
We define
\beq\bsp
\textrm{K}_{\text{QCD}}^{\text{N$^3$LO}}&\,=\frac{\sigma^{(3)}(\mu_f=\mu_r=Q)}{\sigma^{(2)}(\mu_f=\mu_r=Q)}\,,\\
\delta(X)&\,=\frac{\delta_X(\sigma^{(3)})}{\sigma^{(3)}(\mu_f=\mu_r=Q)},
\esp\eeq
where $\sigma^{(n)}(\mu_f=\mu_r=Q)$ is the hadronic cross section including perturbative corrections up to $n^{\text{th}}$ order evaluated for $\mu_F=\mu_R=Q$ and $\delta_X(\sigma^{(n)})$ is the absolute uncertainty of the cross section from source $X$ as described below.
{Furthermore, we show in the last column of tab.~\ref{tab:values} the ratio of the leading order cross section to produce a lepton pair via $Z$ boson and virtual photon exchange~\cite{Campbell:1999ah,Campbell:2011bn,Campbell:2015qma,Campbell:2019dru} over exclusively virtual photon exchange.}

Let us now analyse the two sources of uncertainty related to the PDFs ({PDF+}$\alpha_S$ an PDF-TH) and the dependence of the cross section on the renormalisation and factorisation scales.
Fig.~\ref{fig:pdfs} displays the impact of our imprecise knowledge of parton distribution functions and the strong coupling constant on our abilities to predict the DY cross section. 
{Let us first explain how we evaluate $\delta(\text{PDF+}\alpha_S)$.}
The PDFs and the strong coupling constant cannot be computed from first principle but they need to be extracted from measurements.
In order to study the PDF uncertainties we use the Monte-Carlo replica method following the PDF4LHC recommendation~\cite{Butterworth:2015oua} that uses 100 different PDF sets to compute the 68 \% confidence level interval.
The strong coupling constant uncertainty is computed using two correlated PDF sets provided by ref.~\cite{Butterworth:2015oua} and is then combined in quadrature with the PDF uncertainty to give $\delta(\text{PDF}+\alpha_S)$.
{The uncertainty obtained in this way does not yet include the fact that currently all PDF sets are extracted by comparing experimental to predictions at (at most) NNLO level, nor do they include the next order in the DGLAP equation. A fully consistent N$^3$LO calculation, however, would require the use of a complete set of N$^3$LO PDFs.
We include an uncertainty reflecting the fact that currently there are no N$^3$LO PDF sets available.} 
The estimate of this uncertainty was obtained following the recipe introduced in Ref.~\cite{Anastasiou:2016cez} that uses half the change of the NNLO cross section in changing from NLO to NNLO PDFs as a measure of uncertainty. 
As shown in Fig.~\ref{fig:pdfs} each of the two uncertainties is of the order of $\pm2\%$ over the whole range of invariant masses considered.

 Fig.~\ref{fig:Q_variation} shows the value of the NLO, NNLO and N$^3$LO cross sections normalised to the central N$^3$LO value as a function of the invariant mass $Q^2$ of the lepton pair. 
 The bands indicate the dependence of the cross section at different orders on the choice of the renormalisation and factorisation scales. 
 We choose $Q$ as a central scale and increase and decrease both scales independently by a factor of two  with respect to the central scale  while maintaining $\frac{1}{2}\le \mu_R/\mu_F\le 2$.
 We observe that at N$^3$LO the cross section depends only very mildly on the choice of the scale.
 In particular, for small and very large invariant masses the dependence on the scale is substantially reduced by inclusion of N$^3$LO corrections compared to NNLO. 
 Remarkably, however, we find that for invariant masses $50$ GeV $\lesssim Q\lesssim 400$ GeV, the bands obtained by varying the renormalisation and factorisation scales at NNLO and N$^3$LO do not overlap for the choice of the central scale $Q$ that is conventionally chosen in the literature. 
 This is in stark contrast to the case of the N$^3$LO corrections to the inclusive cross section for Higgs production in gluon and bottom-quark fusion~\cite{Anastasiou:2015ema,Mistlberger:2018etf,Duhr:2019kwi}, where the band obtained at N$^3$LO was always strictly contained in the NNLO band (for reasonable choices of the central scales). 
 We note that this behaviour does not depend on our choice of the central scale, but we observe the same behaviour when the central scale is chosen as $Q/2$.
 Since this is a new feature which has not been observed so far for inclusive N$^3$LO cross section, we analyse it in some detail. 

    \begin{figure*}[!t]
\setlength{\abovecaptionskip}{17.5 pt plus 4pt minus 2pt}
\mbox{
\subfigure{\includegraphics[scale=.24]{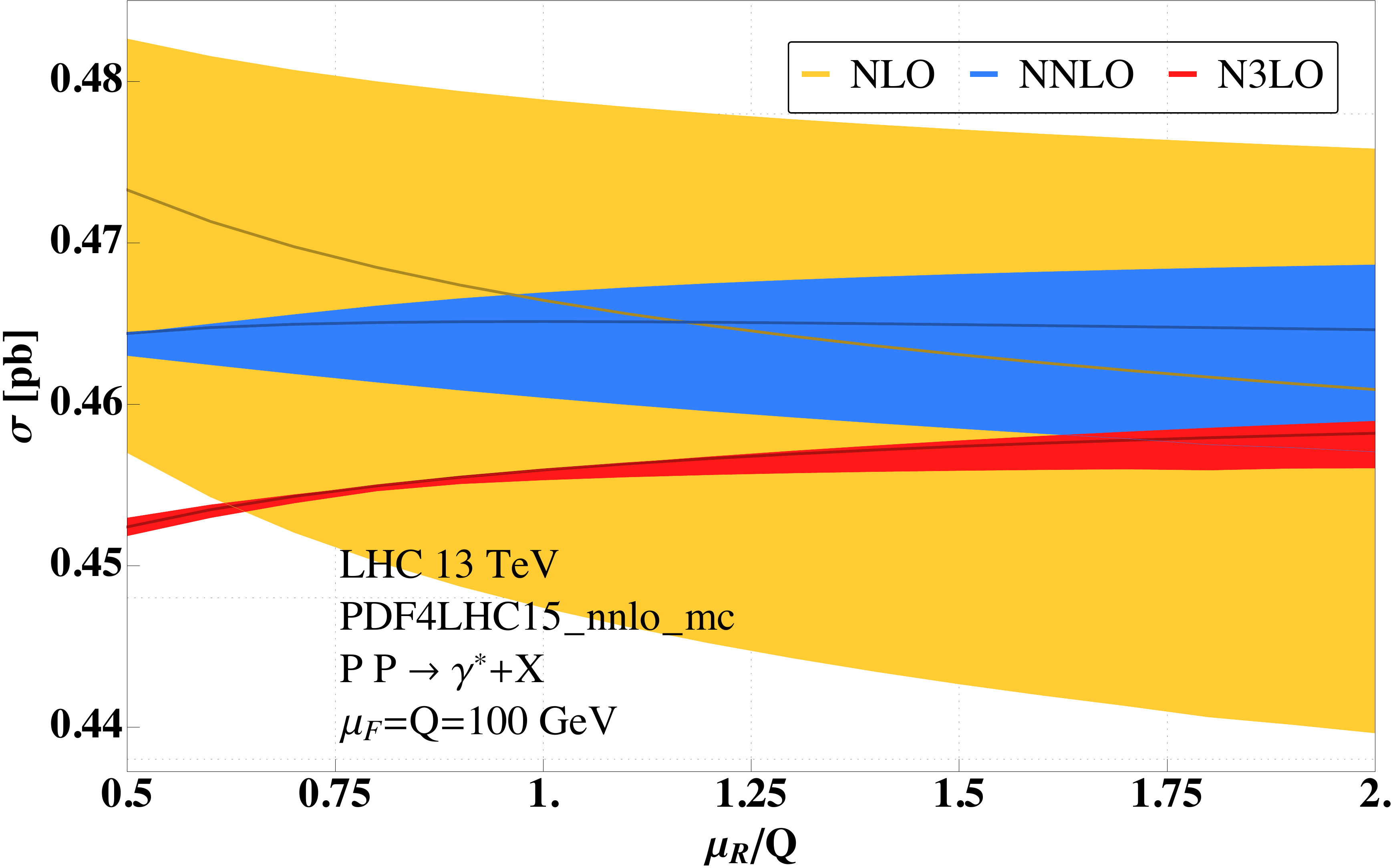}}\hskip 0.5cm
\subfigure{\includegraphics[scale=.24]{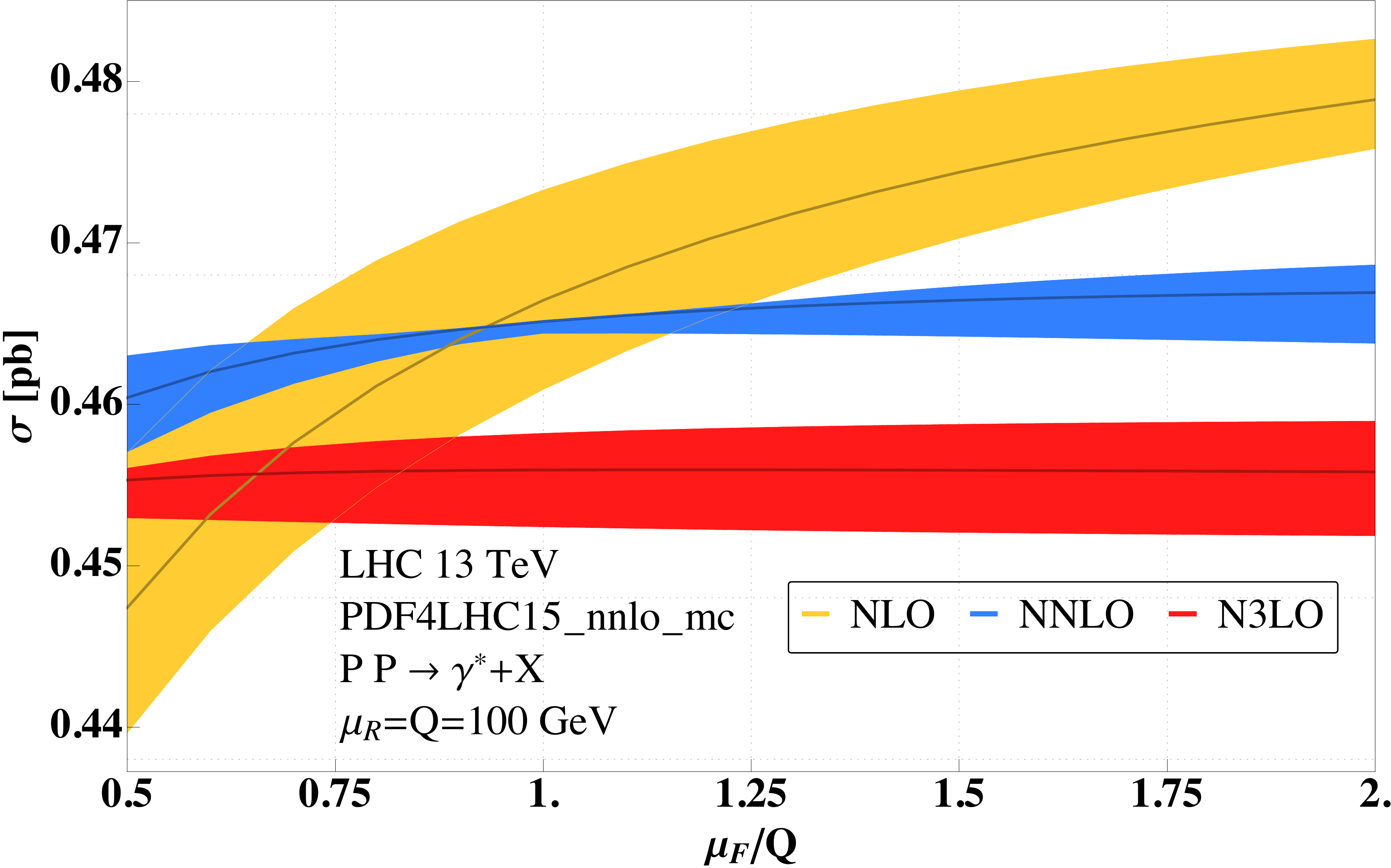}}
}
\vspace{-0.6 cm}
\caption{\label{fig:scales_fixed}Dependence of the cross section on either $\mu_F$ or $\mu_R$ with the other scale held fixed. }
\end{figure*}

Fig.~\ref{fig:scales_fixed} shows the dependence of the cross section for an invariant mass $Q=100$ GeV on one scale with the other held fixed at the central scale $Q=100$ GeV.  
The bands are again obtained by varying the scale by a factor of two up and down around the central scale. 
We see that in both cases the NNLO and N$^3$LO bands do not overlap. 
Furthermore, we see that for the $\mu_R$ dependence the width of the band is substantially reduced when going from NNLO to N$^3$LO. 
For the $\mu_F$ dependence, however, the width of the band is increasing from NNLO to N$^3$LO. 
We note that this statement depends on the choice of the value of $Q^2$ considered as well as the centre-of-mass energy of the hadron collider. {It would be interesting in how far this observation is related to the missing N$^3$LO PDFs (keeping in mind that in that case one could not disentangle completely the PDF-TH and scale uncertainties anymore).}

 \begin{figure}[!b]
  \begin{center}
   \includegraphics[scale=0.25]{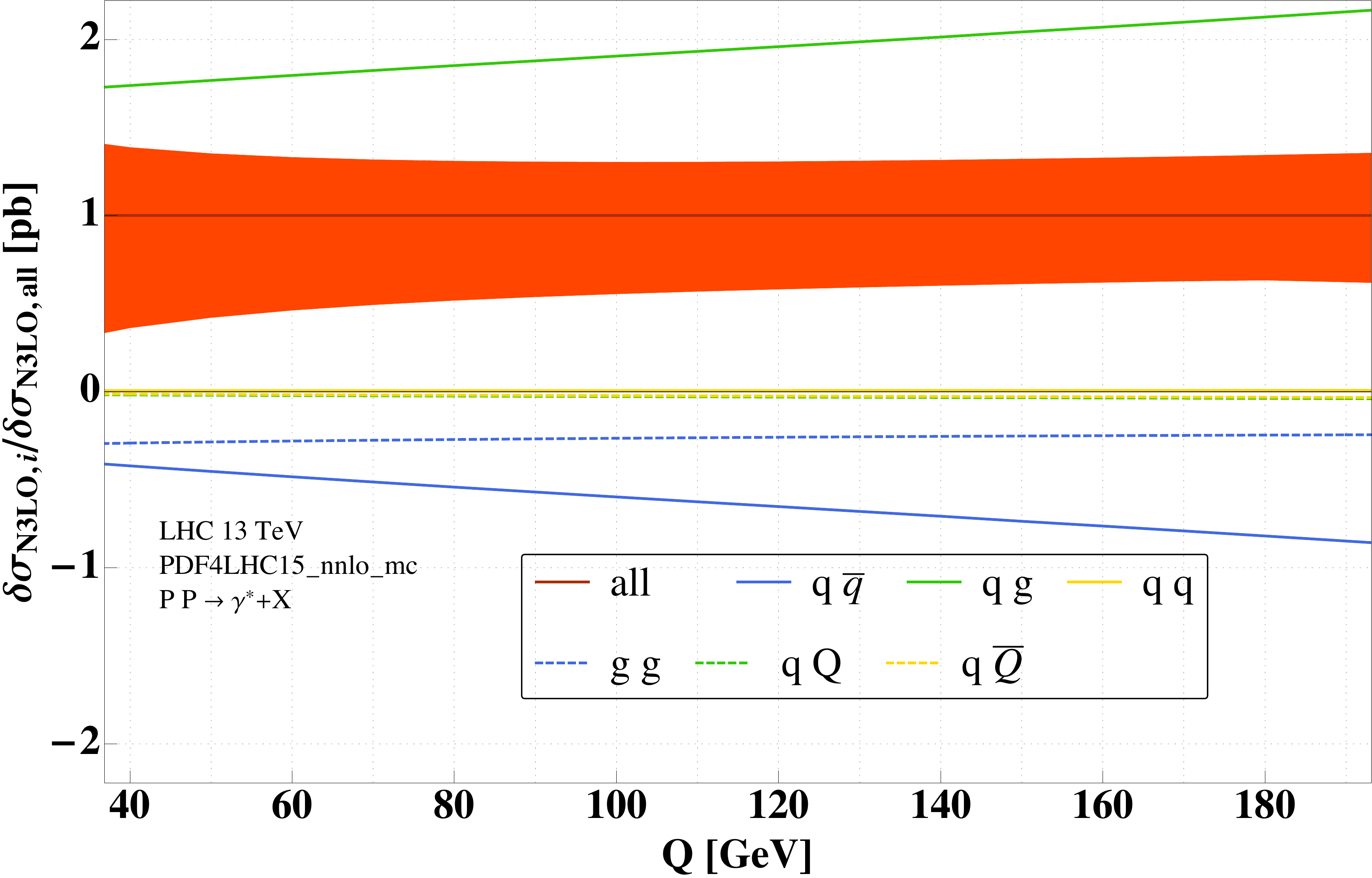}
  \end{center}
  \caption{\label{fig:channels}Contributions of the different partonic channels at N$^3$LO, as a function of the invariant mass $Q$ and normalised to the N$^3$LO cross section. 
  The red band correspond to the variation of the N$^3$LO correction under variation of the factorisation and renormalisation scale as described in the text.
   }
  \end{figure}

Fig.~\ref{fig:channels} shows the relative contribution of the different partonic channels as a function of the invariant mass $Q^2$ to the N$^3$LO correction of the DY cross section. 
We see that the cross section is dominated by the $q\bar{q}$, $qg$ and $gg$ channels. 
While the $qg$ channel gives a large and positive contribution, the $q\bar{q}$ channel (and to a lesser extend also the $gg$ channel) gives a negative contribution which largely cancels the contribution from the $qg$ channel. 
The same cancellation happens already in the case of the NNLO corrections to an even larger extent. 
Given the sizeable cancellation of different partonic initial state contributions, small numerical changes in the parton distribution functions will have an enhanced effect on the prediction of the DY cross section.
Consequently, estimating and improving on the sources of uncertainties related to parton distribution functions considered in Fig.~\ref{fig:pdfs} is of great importance.

 \section{Conclusions}
We have presented for the first time the complete computation of the N$^3$LO corrections in QCD for the production of a lepton pair from a virtual photon. 
Our main findings are percent level corrections to the hadronic cross section and an overall reduction of dependence on the perturbative scales.
The size of this corrections is consistent with N$^3$LO corrections to Higgs boson production in gluon-fusion~\cite{Anastasiou:2015ema,Anastasiou:2016cez,Mistlberger:2018etf} and bottom-quark-fusion~\cite{Duhr:2019kwi} and indicates the importance of N$^3$LO corrections to LHC processes for phenomenology conducted at the percent level.
 
 In the region of small invariant masses where the contribution from the $Z$ boson is small, $Q\lesssim 50$ GeV, the photon contribution computed here is the dominant part of the cross section.
 For other kinematic regions we expect the K-factor of the $Z$ boson contribution to behave qualitatively very similarly to the photon contribution and our results provide essential information.
 We see from Fig.~\ref{fig:Q_variation} that our computation substantially reduces the dependence of the cross section on the renormalisation and factorisation scales.  
 In contrast to the corrections to Higgs boson production, however, the shift of the predicted value of the DY cross section due to the inclusion of N$^3$LO corrections is not contained in the naive scale variation bands of NNLO predictions for all values of $Q$.
 We emphasise that this should not be interpreted as an indication of a breakdown of perturbative QCD, but rather as a sign that uncertainty estimates based on a purely conventional variation of the scales should be taken with a grain of salt. 
 Moreover, we observe an intricate pattern of large cancellations of contributions from different partonic initial states at NNLO and N$^3$LO.
 This implies a large sensitivity of the cross section on relatively small shifts in parton distribution functions. 
 In combination with the fact that the DY process is a key ingredient for the determination of PDFs, this motivates to push for parton distributions determined from N$^3$LO cross sections in the future.
 It also hints at am intricate entanglement of PDFs and the structure of QCD cross sections, so that the uncertainty estimate obtained from scale variation cannot be completely disentangled from the PDF-TH uncertainties. The perturbative uncertainty should rather be seen as the combination of the two.
 Finally, we believe that our findings warrant a critical revision of the strategy to assess perturbative uncertainties and the consequences thereof on PDF determination etc.

\begin{acknowledgments}
\emph{Acknowledgements:}  
We are grateful to Babis Anastasiou, Lance Dixon, Stefano Forte and Simone Marzani for discussion.
The research of CD is supported by the ERC grant 637019 ``MathAm''. 
BM is supported by the Pappalardo fellowship, and the research of FD is supported by the  U.S.  Department  of Energy (DOE) under contract DE-AC02-76SF00515.
\end{acknowledgments}

\bibliography{main}

\end{document}